\documentclass[twocolumn, secnumarabic, amssymb, nobibnotes, aps]{revtex4-1}
\usepackage{graphicx}

\begin{document}

\title{\textit{Ab initio} studies of Co$_2$FeAl$_{1-x}$Si$_x$ Heusler alloys}
\author{N. Gonzalez Szwacki}
\email{gonz@fuw.edu.pl}
\author{Jacek A. Majewski}
\affiliation{Institute of Theoretical Physics, Faculty of Physics, University of Warsaw, ul. Pasteura 5, PL-02-093 Warsaw, Poland}
\date{12.02.15}

\begin{abstract}
We present results of extensive theoretical studies of Co$_2$FeAl$_{1-x}$Si$_x$ Heusler alloys, which have been performed in the framework of density functional theory employing the all-electron full-potential linearized augmented plane-wave scheme. It is shown that the Si-rich alloys are more resistive to structural disorder and as a consequence Si stabilizes the $L2_1$ structure. Si alloying changes position of the Fermi level, pushing it into the gap of the minority spin-band. It is also shown that the hyperfine field on Co nuclei increases with the Si concentration, and this increase originates mostly from the changes in the electronic density of the valence electrons.
\end{abstract}


\maketitle

\section{Introduction}
Heusler alloys are usually ternary X$_2$YZ compounds that consist of two transition metals X (e.g., Co, Ni, Cu, Pd) and Y (e.g., Ti, Mn, Fe, Zr, Hf), and one main group element Z (e.g., Al, Si, Ga, In, Sn, Sb) and crystallize in the $L2_1$ structure with the \textit{Fm$\bar{3}$m} space group \cite{a1}. They have attracted considerable research attention in recent years due to their promising applications in magneto-electronic and spintronic devices \cite{a1,a2}. In particular, Heusler compounds based on cobalt behave like half-metallic ferromagnets, i.e., the electronic band calculations for these compounds show a metallic density of state (DOS) only for the spin direction of the majority band and a band gap around the Fermi level for the other spin direction of the minority band \cite{a2}. Co$_2$YZ alloys have a compatible lattice structure with the industrially used zincblende semiconductors and possess a high Curie temperature ranging between 690 and 1100 K that allows the applications in the devices operating at room temperature \cite{a2}. Those materials exhibit also large values of the magnetic moment, e.g. 5 and 6~$\mu$$_B$ for Co$_2$FeAl and Co$_2$FeSi, respectively \cite{a2}, and their magnetic properties are governed by the localized \textit{d} electrons interacting via spin-polarized itinerant electrons \cite{a3}. Defects and antisite disorder play a crucial role among the factors destroying the half-metallic properties, and therefore reducing spin polarization \cite{a4}. These alloys can transform from the $L2_1$ structure into \textit{B}2 upon disorder between Y and Z atoms or into \textit{A}2 upon disorder between X, Y and Z atoms \cite{a2}.

In this paper, we report results of first principles calculations for Co$_2$FeAl$_{1-x}$Si$_x$ (\textit{x}= 0, 0.25, 0.5, 0.75, 1) Heusler alloys. These alloys have attracted considerable experimental \cite{a2,a5,a6,a9} and theoretical \cite{a6,a7} attention. The theoretical investigations were focused mainly on band structure calculations using several computational approaches. Here, apart from calculating the electronic and magnetic properties of the alloys, we investigate the magnetic hyperfine fields at Co nuclei for the ordered $L2_1$ structures. We also present results of total energy calculations for alloys having the \textit{B}2 structure, namely where positions of Fe atoms and atoms at the Z sites (i.e., Al or Si) are swapped. This allows for the calculation of the energy difference, $\Delta E$, between the structure containing the antisite defect and the highly ordered $L2_1$ structure, and further for the estimation of the defect occurrence for a given concentration \textit{x} in the studied Co$_2$FeAl$_{1-x}$Si$_x$ Heusler alloys.

The rest of the paper is organized as follows. In Sec.~II the computational details employed in the present calculations are discussed. In Sec.~III the results are presented, beginning with a discussion in Sec.~IIIA on how the parameter \textit{U} is chosen, what is an important issue in the GGA+\textit{U} method. The results for ordered $L2_1$ structures are presented in Sec.~IIIB followed by Sec.~IIIC dedicated to the results for structures with antisite defects. We then end with some concluding remarks in Sec.~IV.

\section{Computational approach}
The computations are performed using the Elk code \cite{a8}, implementing the all-electron full-potential linearized augmented plane-wave (FLAPW) method. Various exchange-correlation functionals (PBE, PBEsol, and WC06) are tested within the GGA+\textit{U} approximation. The same \textit{U} parameter is used for both Fe and Co atoms and the intra-atomic exchange parameter, \textit{J}, is set to zero. The muffin-tin radii $R_{MT}$ are set to 2.0~a.u. for all atoms. The plane-waves cutoff is set to $R_{MT}\times G_{max}=8.0$ and a \textbf{k}-point mesh of $20 \times 20 \times 20$ is used. The angular momentum expansion in the muffin tin spheres is taken up to $l_{max}=7$ for both the potential and density. The maximum \textbf{G}-vector length for the density and potential expansion in the interstitial region is set to $12~a.u.^{-1}$. The spin-orbit coupling is neglected and the experimental lattice parameters (5.73 and 5.64~\AA\ for Co$_2$FeAl and Co$_2$FeSi, respectively \cite{a2}) are used in the calculations. To study the alloying effects, we use a supercell containing 16 atoms (4 times the primitive unit cell of Co$_2$FeAl or Co$_2$FeSi) with 4 Z sites, which are occupied by Al and Si atoms, with Si concentration, \textit{x}, equal to 0.0, 0.25, 0.5, 0.75, and 1.0. The lattice constants for the intermediate compositions of Co$_2$FeAl$_{1-x}$Si$_x$ are estimated from the Vegard's law. For each configuration, we perform optimization of the atomic positions.

\section{Results}
\subsection{The choice of \textit{U}}

\begin{figure} [b]
\centering
\includegraphics[width=8.6 cm]{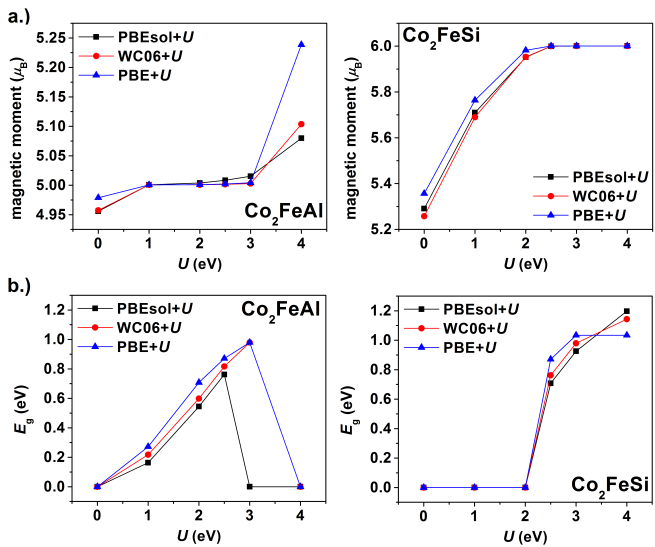}
\caption{(color online). (a) Magnetic moment dependence on the choice of the \textit{U} parameter for Co$_2$FeAl (left) and Co$_2$FeSi (right). (b) Energy gap, $E_g$, in the minority band vs. the value of the \textit{U} parameter for Co$_2$FeAl (left) and Co$_2$FeSi (right).}
\label{fig1}
\end{figure}

Some properties of Co$_2$FeAl and Co$_2$FeSi alloys have been already presented in previous theoretical works (see Refs. 6,8 and references therein). It is known, for instance, that it is necessary to include the on-site correlation, \textit{U}, in the calculations for Co$_2$FeSi in order to have the half-metallic ground state. Our first step, therefore, was to calibrate the \textit{U} parameter in order to obtain experimental \cite{a2} results for the magnetic moment and electronic properties of Co$_2$FeAl and Co$_2$FeSi. As it is seen in Fig.~\ref{fig1}a~(left), the magnetic moment of Co$_2$FeAl is close (or equal) to 5$\mu$$_B$ for all \textit{U} values up to $\sim$3~eV. Also for \textit{U} values up to $\sim$3~eV ($\sim$2.5~eV for the PBEsol+\textit{U} approach) a nonzero minority energy gap is present for this alloy as it is shown in Fig.~\ref{fig1}b~(left). On the other hand, the half metallic character of Co$_2$FeSi is reproduced only for certain values of \textit{U}. This is shown in Fig.~\ref{fig1}b~(right) where the minority energy gap opens up for values of \textit{U} exceeding 2.5~eV. For those values, the magnetic moment is very close to 6$\mu$$_B$ as it is shown in Fig.~\ref{fig1}a~(right). It is worth mentioning that the above-described picture is almost independent on the exchange-correlation functional that is used (PBE, WC06 or PBEsol). Therefore, from now on, we will present results only for the PBE exchange-correlation functional and the \textit{U} parameter set to 2.5~eV. As it turns out, this choice of \textit{U} and the fuctional reproduces well the properties of the end members of Co$_2$FeAl$_{1-x}$Si$_x$.
\subsection{Ordered $L2_1$ structures}

\begin{figure} [b]
\centering
\includegraphics[width=7.5 cm]{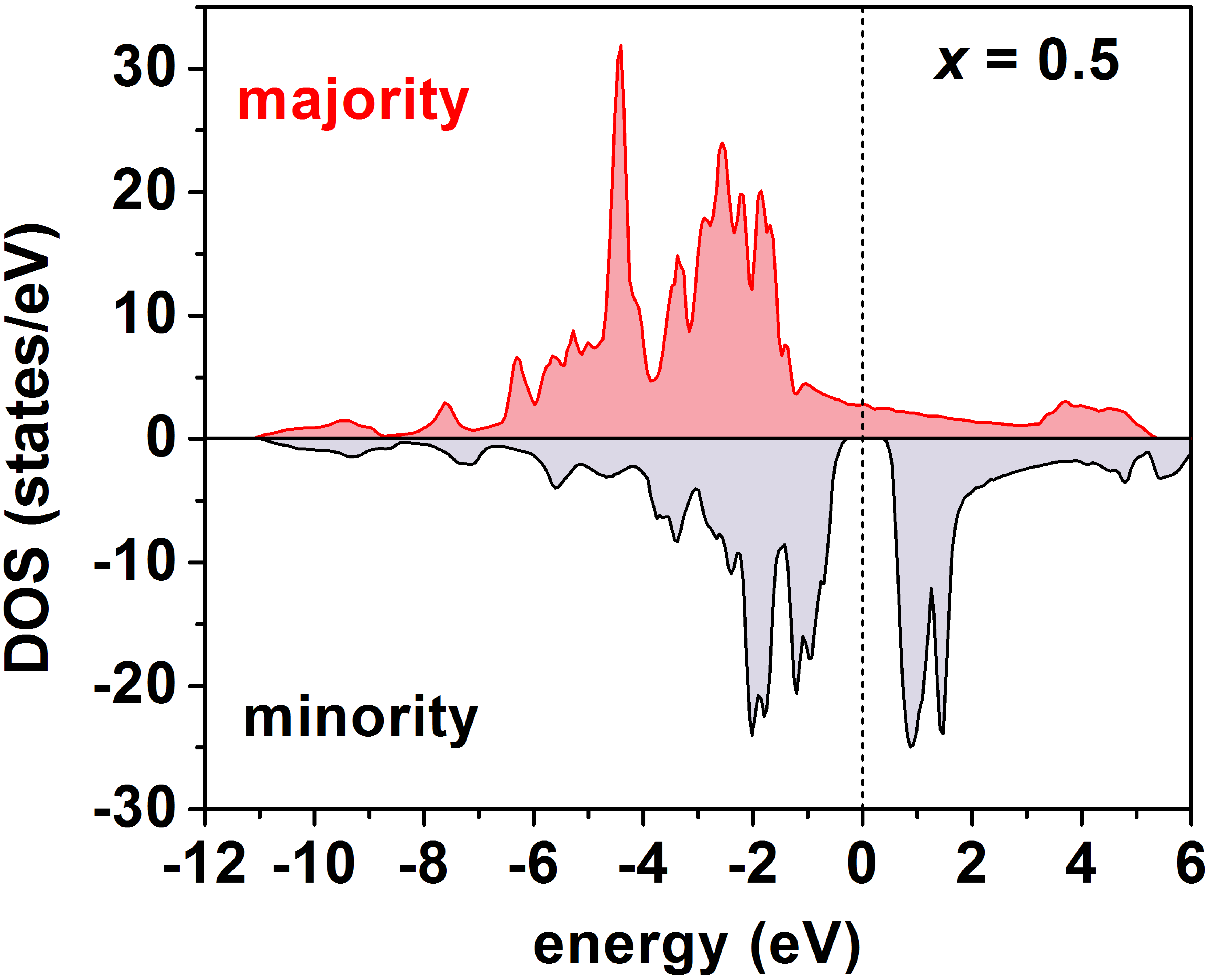}
\caption{(color online). Spin resolved density of states of Co$_2$FeAl$_{1-x}$Si$_x$ for \textit{x}=~0.5. The positive and negative DOS values represent spin-up (majority) and spin-down (minority) bands, respectively. The Fermi level is set to 0~eV as shown by the dashed line.}
\label{fig2}
\end{figure}
Our band structure calculations for Co$_2$FeAl$_{1-x}$Si$_x$ show for each concentration \textit{x} a finite DOS at the Fermi level only for the spin direction of the majority band and a band gap for the spin direction of the minority band. For Co$_2$FeAl the Fermi level lies just above the valence band, whereas for Co$_2$FeSi lies just below the conduction band. This result is in accord with previous reports \cite{a6,a7} that also show a shift of the majority spin density that is compensated by a shift of the minority spin density with increasing Si content, and as a result a virtual movement of the Fermi energy in the minority gap is observed. Figure~\ref{fig2} shows the spin resolved density of states for Co$_2$FeAl$_{0.5}$Si$_{0.5}$. For that composition, calculations predict that the Fermi level is located in the middle of the band gap of the minority band. Such position of the Fermi level enhances the thermal stability of the half-metallicity because reduces the sensitivity to thermal fluctuation effects \cite{a13}.

The magnetic properties of Co$_2$FeAl$_{1-x}$Si$_x$ can be summarized as follows. The end alloys Co$_2$FeAl and Co$_2$FeSi have integer magnetic moments (5 and 6~$\mu$$_B$, respectively) that obey the generalized Slater-Pauling rule \cite{a14} ($\textit{m}=(\textit{N}_v-24)\mu_B$, where $\textit{N}_v$ is the number of valence electrons in the primitive cell). The substitution of Al by Si results in a fractional magnetic moment that varies linearly with \textit{x} ($\textit{m}_x=(5+\textit{x})\mu_B$). Each Si contributes with one additional electron that fills the majority band and increases the localized magnetic moments on Co and Fe.

\begin{figure} [t]
\centering
\includegraphics[width=7.5 cm]{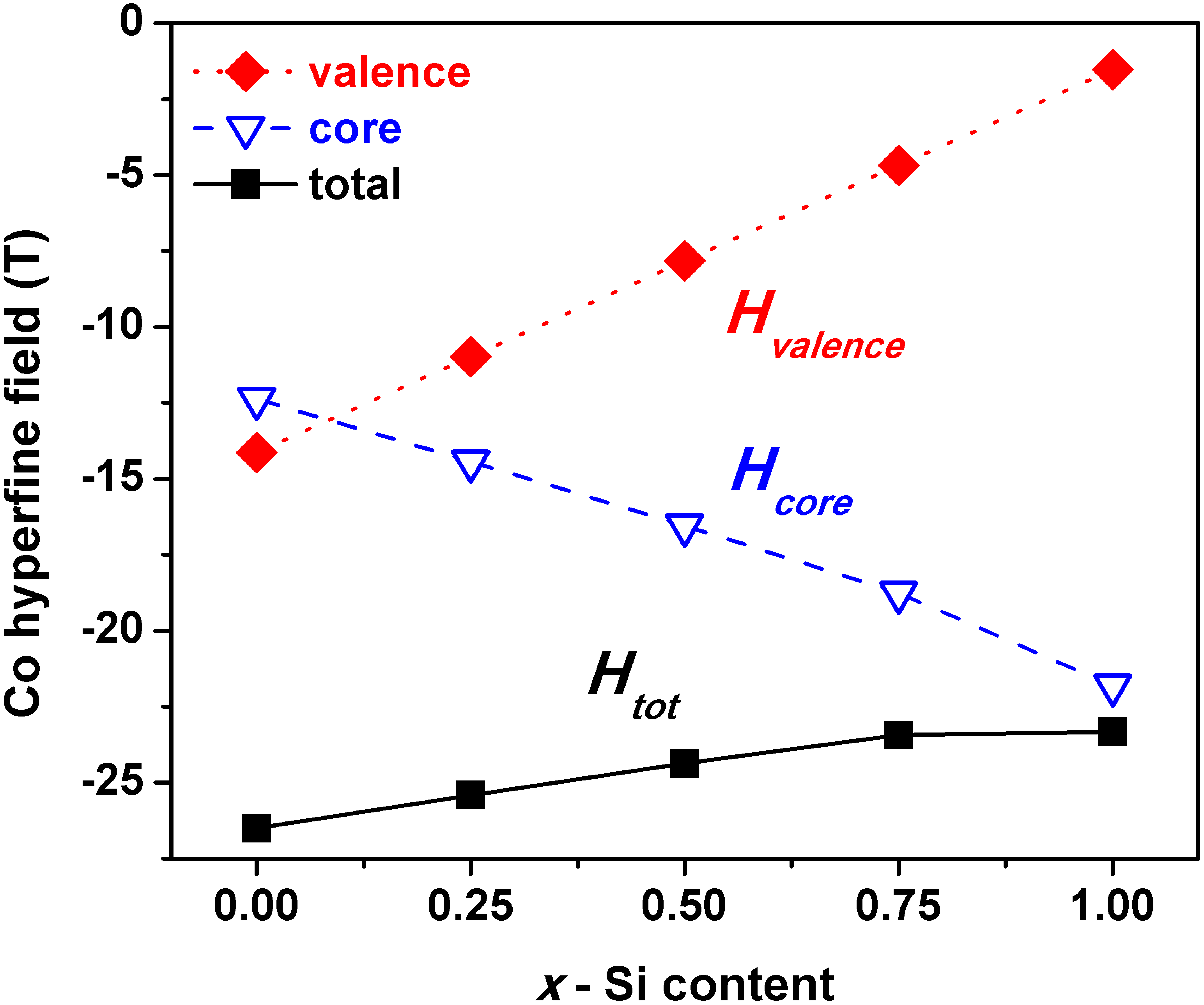}
\caption{(color online). Co hyperfine field ($H_{tot}$) for Co$_2$FeAl$_{1-x}$Si$_x$ as a function of the composition. The 4\textit{s} valence electrons contribution ($H_{valence}$) and the core electrons contribution ($H_{core}$) to the Co hyperfine field is also included.}
\label{fig3}
\end{figure}

Turning now to the discussion of magnetic hyperfine fields in the Co$_2$FeAl$_{1-x}$Si$_x$ Heusler alloys, it has been shown experimentally \cite{a5} that the hyperfine field on Co nuclei increases with the Si concentration, and this increase originates mostly from the changes in the electronic density of the valence electrons. To study this, we have calculated the onsite hyperfine field on the Co atoms of Co$_2$FeAl$_{1-x}$Si$_x$ as a function of the composition. The onsite hyperfine field which for ferromagnetic 3\textit{d} metals consist mainly of the Fermi contact term \cite{a10,a11} is plotted in Fig.~\ref{fig3}. The hyperfine field ($H_{tot}$) is often regarded \cite{a12} as a sum of two separate contributions, $H_{core}$ and $H_{valence}$, originating from the polarization of core (1\textit{s}, 2\textit{s}, 3\textit{s}) and valence (4\textit{s}) electrons, respectively, and those contributions are also reflected in Fig.~\ref{fig3}. The qualitative agreement between experimental results \cite{a5} and those reported here is very good. The slope of $H_{tot}$ as a function of composition is well reproduced, however the values of the hyperfine field are shifted down with respect to experimental values. This discrepancy between theory and experiment could arise on one hand from the non-sufficient description (in GGA+\textit{U} calculations) of the effect of correlation on the 3\textit{d}-like electron wave functions \cite{a10}, and on the other hand from the fact that orbital and dipolar contributions to the hyperfine field are neglected in our calculations. The positive slope of the valence electrons contribution (see Fig.~\ref{fig3}) to the hyperfine field means that the population of 4\textit{s} electrons with spin up increases faster with Si content than the population of electrons with spin down. This is consistent with other theoretical predictions \cite{a7} showing that, upon increase of the Si content, the electrons dope the majority-spin band, leading to a virtual shift of the Fermi level inside the gap of the minority-spin band.

\subsection{Structures with antisite defects}
\begin{figure} [t]
\centering
\includegraphics[width=7.5 cm]{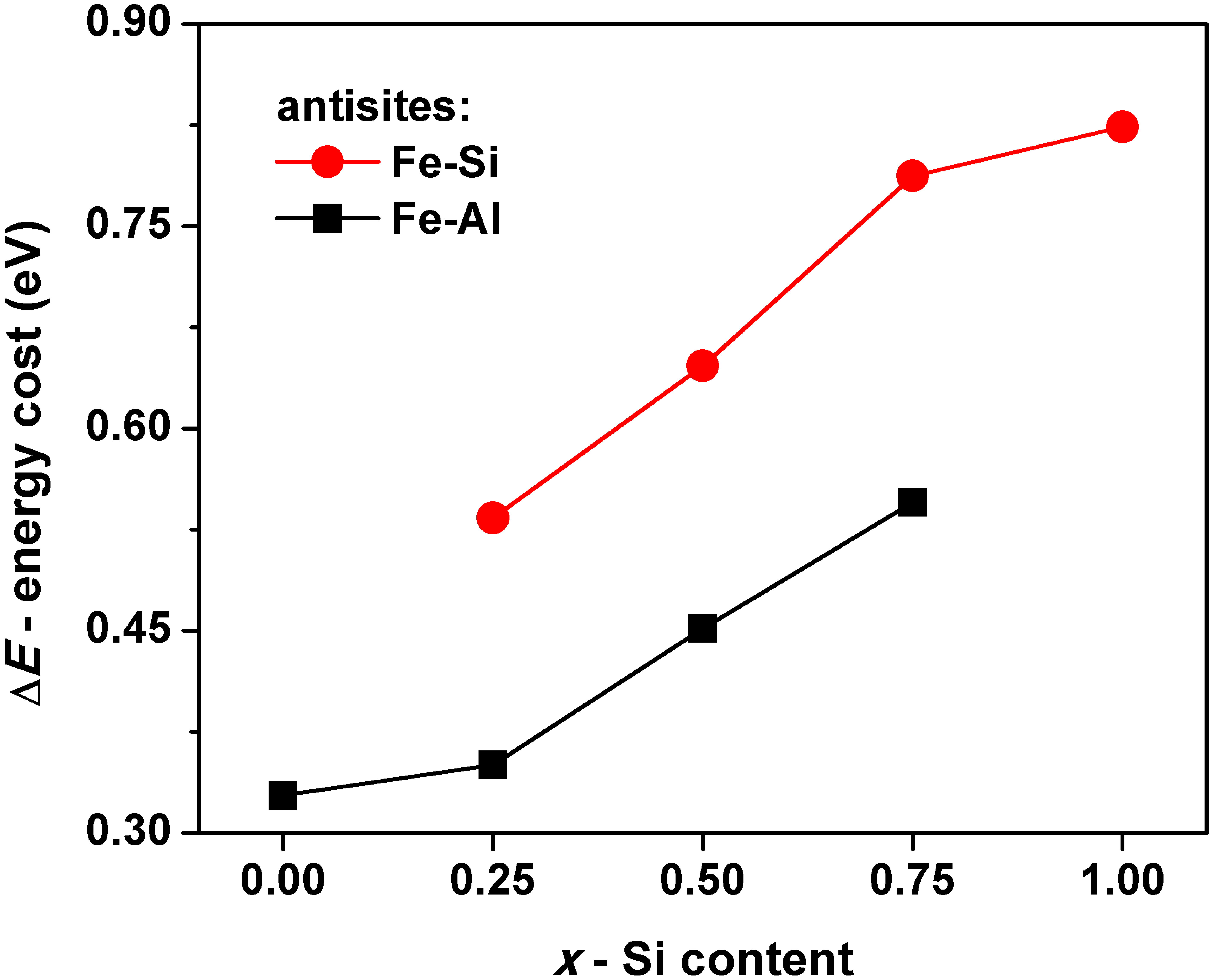}
\caption{(color online). Energy cost for the formation of Fe-Al and Fe-Si antisites in Co$_2$FeAl$_{1-x}$Si$_x$ as a function of the composition. $\Delta E$ is defined as the total energy difference of Co$_2$FeAl$_{1-x}$Si$_x$ with and without the defect.}
\label{fig4}
\end{figure}

As mentioned above the Heusler alloys appear not only in an ordered structure but also exhibit pronounced alloying. Besides the ordered $L2_1$ structure, a commonly occurring disordered alloy have the \textit{B}2 structure. It is known that even a small mixture of atoms in Y and Z positions may affect the electronic structure and magnetic properties of the Heusler alloys \cite{a6}. Our calculations demonstrate, in accordance with experiment \cite{a5}, that the probability of creation of Fe-Al antisites decreases with the growing Si concentration. This is shown in Fig.~\ref{fig4}, where we plotted the energy cost, $\Delta E$, for the formation of Fe-Al and also Fe-Si antisites as a function of the Si composition. $\Delta E$ clearly increases with \textit{x} both for Fe-Al and Fe-Si antisites. From Fig.~\ref{fig4} it is also clear that the formation of Fe-Si antisites is less probable than that of Fe-Al antisites, what helps to explain the luck of satellite structures in the $^{59}$Co NMR spectra for Co$_2$FeSi as oppose to the case of Co$_2$FeAl \cite{a5}.

\section{Conclusions}
In conclusion, our theoretical calculations show that the hyperfine field on Co nuclei increases with the Si concentration, and this increase originates mostly from the changes in the electronic density of the valence electrons. The Si doping is, therefore, responsible for the shift of the Fermi energy in the energy gap of the minority band. Our study also corroborates that the stronger \textit{p}-\textit{d} hybridization in Si-rich than Al-rich alloys, makes the Si-rich alloys more resistive to structural disorder and as a consequence Si stabilizes the desired $L2_1$ structure. Our theoretical findings agree fairly well with available experimental data. 

\begin{acknowledgements}
This paper has been supported by the project No. N N507 453138 of Ministry of Science and Higher Education. We also acknowledge the Interdisciplinary Center of Modeling at the University of Warsaw for the access to computer facilities.
\end{acknowledgements}

\bibliographystyle{apsrev4-1} 
\bibliography{v3}

\end{document}